\documentclass[journal,transmag]{IEEEtran}
\usepackage{cite,graphicx,amssymb,amsmath,psfrag,bm}
\usepackage[usenames,dvipsnames,table,xcdraw]{xcolor}
\usepackage{MnSymbol}
\usepackage{array}
\usepackage{hyperref}
\usepackage{cancel}
\usepackage{epstopdf}
\usepackage{rotating}
\usepackage{pstool}
\usepackage{multirow}
\usepackage{color}
\usepackage{soul}
\usepackage{fullpage}
\usepackage{bibentry}
\usepackage{caption}
\usepackage{subcaption}

\nobibliography*

\newcommand{\ve}[1]{\boldsymbol{#1}}
\newcommand{\te}[1]{\overline{\overline{#1}}}

\begin{document}
\title{Computational Analysis of Metasurfaces}


\author{\IEEEauthorblockN{Yousef Vahabzadeh,
Nima Chamanara,
Karim Achouri and
Christophe Caloz~\IEEEmembership{Fellow,~IEEE},}
\IEEEauthorblockA{Polytechnique Montr\'{e}al, QC, Canada}
\thanks{Corresponding author: Y. Vahabzadeh (email: yousef.vahabzadeh@polymtl.ca).}}

\markboth{JMMCT}%
{Shell \MakeLowercase{\textit{et al.}}: Bare Demo of IEEEtran.cls for IEEE Transactions on Magnetics Journals}

\IEEEtitleabstractindextext{%
\begin{abstract}
Metasurfaces represent one of the most vibrant fields of modern science and technology. A metasurface is a complex electromagnetic structure, that is typically deeply subwavelength in thickness, electrically large in transverse size and composed of subwavelength scattering particles with extremely small features; it may generally be bianisotropic, space-varying and time-varying, nonlinear, curved and multiphysics. With such complexity, the design of a metasurface requires a holistic approach, involving synergistic synthesis and analysis operations, based on a solid model. The Generalized Sheet Transition Conditions (GSTCs), combined with bianisotropic surface susceptibility functions, provide such a model, and allow now for the design of sophisticated metasurfaces, which still represented a major challenge a couple of years ago. This paper presents this problematic, focusing on the \emph{computational analysis} of metasurfaces via the GSTC-susceptibility approach. It shows that this analysis plays a crucial role in the holistic design of metasurfaces, and overviews recently reported related frequency-domain (FDFD, SD-IE, FEM) and time-domain (FDTD) computational techniques.
\end{abstract}

}

\maketitle

\IEEEdisplaynontitleabstractindextext

\IEEEpeerreviewmaketitle

\section{Introduction}\label{Sec:Intro}

Metasurfaces are two-dimensional arrays of subwavelength metallic or dielectric scattering particles that transform electromagnetic waves in various ways~\cite{Kuester, Holloway_Metafilm_RT, Capasso_Falt_optics, Christopher_Metafil_mts_character, Holloway_mts_overview, Glybovski_mts_uwave_visible, Urbas_2016_Metamat_roadmap, Minovich_2015_functional_nonlin_optic_ms}. Compared to 3D metamaterials~\cite{Metamaterials2009}, they are less lossy, easier to fabricate, and offer a broader range of functionalities. Within less than a decade, they have already led to a myriad of applications, such as for instance flat lenses~\cite{Pors_broadband_Plasmon_mts, Ma_chiral_mts}, vortex wave generators~\cite{VVBeam,KA_mts_princ_illust}, nonlinear beam shapers~\cite{keren2015nonlinear,KA_2nd_order_mts} and remote processors~\cite{KA_mts_spatial_Proces}. Moreover, they offer the possibility to simultaneously perform multiple independent field transformations~\cite{taravati2017nonreciprocal, Karim_general_mts_syn, achouri_2017_AngSC, achouri2015birefringent}.

\begin{figure}[!ht]
\centering
\includegraphics[width=1\columnwidth]{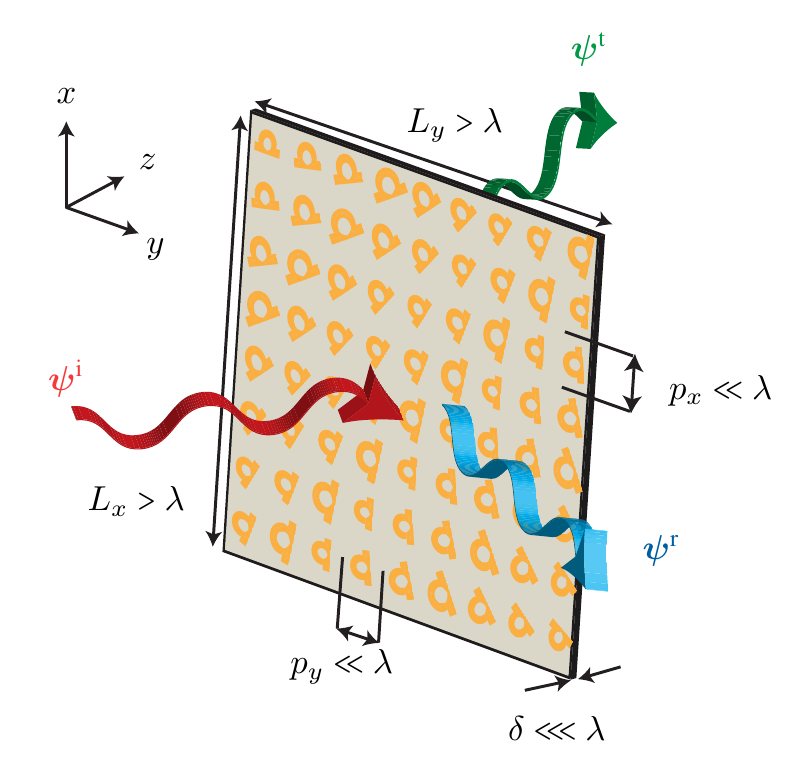}
\caption{General metasurface, transforming an incident wave $\boldsymbol{\psi}^\textrm{i}$ into a reflected wave $\boldsymbol{\psi}^\textrm{r}$ and a transmitted wave $\boldsymbol{\psi}^\textrm{t}$.}
\label{Fig:Metasurface}
\end{figure}

\begin{figure}[!ht]
\centering
\includegraphics[width=1\columnwidth]{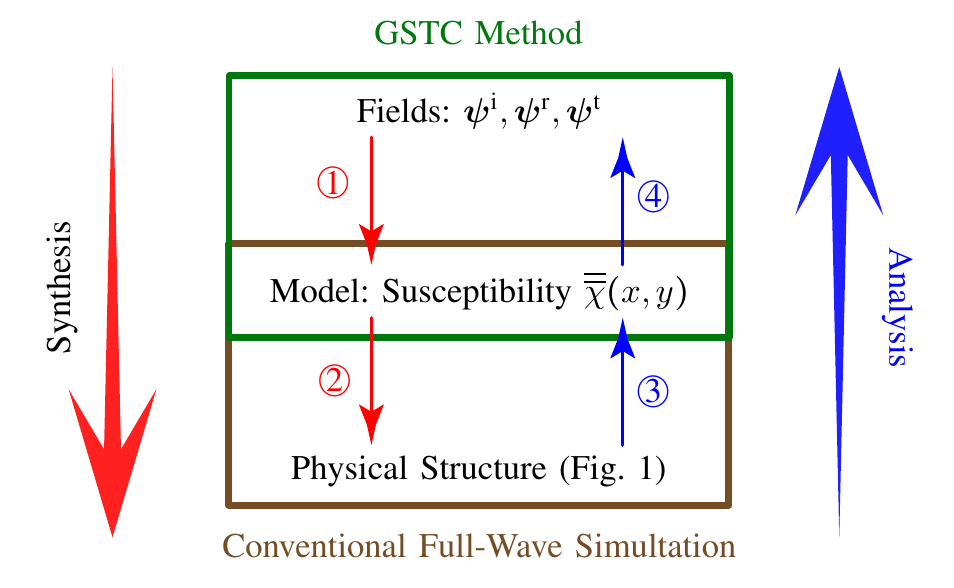}
  \caption{Holistic design of a metasurface.}
  \label{Fig:Design}
\end{figure}

Figure~\ref{Fig:Metasurface} represents a general metasurface, transforming an incident wave $\boldsymbol{\psi}^\textrm{i}$ into a reflected wave $\boldsymbol{\psi}^\textrm{r}$ and a transmitted wave $\boldsymbol{\psi}^\textrm{t}$. As suggested in the figure, a metasurface is typically an electrically very thin ($\delta\lll\lambda$), electrically relatively large ($L_x,L_y>\lambda$), homogenizable ($p_x,p_y\ll\lambda$) non-periodic and bianisotropic~\cite{Cloud_Rothwell} electromagnetic structure. It is therefore challenging to model and typically requires a holistic design approach of the type described in Fig.~\ref{Fig:Design}, which includes both synthesis and analysis operations.

The synthesis operation globally consists in determining the physical (geometrical and electromagnetic) parameters of the metasurface structure, such as the metasurface size and scattering particle geometries, to achieve a specified wave transformation~\cite{Pfeiffer_Bianiso_mts,Karim_general_mts_syn,Holloway_Metafilm_RT}. It may be decomposed in the following two operations: the determination of a continuous tensorial susceptibility function, $\te{\chi}(x,y)$, characterizing the metasurface from the specified fields [$\raisebox{.5pt}{\textcircled{\raisebox{-.9pt} {1}}}$ in Fig.~\ref{Fig:Design}] and, based on this function, the determination of the aforementioned physical parameters [$\raisebox{.5pt}{\textcircled{\raisebox{-.9pt} {2}}}$ in Fig.~\ref{Fig:Design}]. The analysis operation is the inverse of the synthesis. It consists in determining the scattered fields of a given physical metasurface, and may be itself decomposed in the following two operations: the extraction of the susceptibility function $\te{\chi}(x,y)$ characterizing the metasurface structure [$\raisebox{.5pt}{\textcircled{\raisebox{-.9pt} {3}}}$ in Fig.~\ref{Fig:Design}] and the subsequent computation of the scattered fields [$\raisebox{0.8pt}{\textcircled{\raisebox{-.9pt} {4}}}$ in Fig.~\ref{Fig:Design}]~\cite{GSTC_FDFD,GSTC_FDTD}. This paper focusses on the last [$\raisebox{0.8pt}{\textcircled{\raisebox{-.9pt} {4}}}$] of these operations, whose importance is tremendous both for the analysis itself and for the sysnthesis as will be explained in Sec.~\ref{sec:usefulness_GSTC_Analysis}.

Being typically much thinner than the operating wavelength, a metasurface is most efficiently modeled as a sheet of \emph{zero thickness} ($\delta=0$). Moreover, it may support both electric and magnetic field discontinuities and exhibit arbitrary bianisotropy. Consequently, a metasurface may be modeled by the four $3\times 3$ (possibly space- and time-varying) \emph{surface} susceptibility tensors $\bar{\bar{\mathbf{\chi}}}_\text{ee}$, $\bar{\bar{\mathbf{\chi}}}_\text{mm}$, $\bar{\bar{\mathbf{\chi}}}_\text{me}$ and $\bar{\bar{\mathbf{\chi}}}_\text{em}$, which correspond to the bulk medium parameters $\bar{\bar{\mathbf{\epsilon}}}$, $\bar{\bar{\mathbf{\mu}}}$, $\bar{\bar{\mathbf{\zeta}}}$ and $\bar{\bar{\mathbf{\xi}}}$, respectively~\cite{Cloud_Rothwell}. Such a complex zero-thickness sheet cannot be modeled by conventional boundary conditions~\cite{Volakis_Approx_BDs}, and no currently existing commercial software can efficiently simulate it.

An appropriate approach to model a metasurface tensorial surface susceptibility sheet is the \emph{Generalized Sheet Transition Conditions (GSTCs)}\footnote{The term `GSTCs' seems to first appear in~\cite{Volakis_GSTC} in the context of the modeling of multilayer planar structures.}~\cite{Idemen,Karim_general_mts_syn,Kuester}. This approach generally describes the metasurface discontinuity in terms of an expansion over derivatives of the Dirac delta distribution. The characterization may equivalently be done in terms of polarizability~\cite{Tretyakov_Polarzibility_mts, Tretyakov_absorber_mts} or impedance~\cite{Pfeiffer_Mts_3layer, WONG_synth_Huy_MTS} tensors.

The GSTCs were initially applied to synthesis problems~\cite{Karim_general_mts_syn, Kuester}. Their first application to analysis problems was done in the Finite-Difference Frequency-Domain (FDFD) scheme~\cite{GSTC_FDFD_APS,GSTC_FDFD}. This scheme can efficiently handle dispersive and bianisotropic metasurfaces. However, as a frequency-domain technique, it is inherently monochromatic, and therefore improper for transient analysis, slow in broadband problems and inapplicable to nonlinear problems. Moreover, it requires large memory resources. A GSTC Spectral-Domain Integral-Equation (SD-IE) was reported in~\cite{NC_SD_IE} to speed up the simulation time and reduce the memory requirement of the FDFD implementation. However, this technique introduces other limitations, such as the impossibility to incorporate scattering objects in the computational region, and the difficulty to handle bianisotropic metasurfaces. In~\cite{GSTC_FEM}, the GSTCs were adapted to the Finite Element Method (FEM), which is known to be most efficient for commercial softwares, but which is also tedious to implement while still suffering from the fundamental limitations of frequency-domain techniques.

In order to access the fundamental benefits of time-domain techniques, the GSTCs have been implemented in several Finite-Difference Time-Domain (FDTD) schemes~\cite{Shulabh_FDTD2, Shulabh_FDTD3, Shulabh_FDTD1, GSTC_FDTD}. The GSTC-FDTD proposed in~\cite{GSTC_FDTD} is straightforwardly integrable in existing FDTD codes and can simulate arbitrary space-time varying bianisotropic metasurfaces. However, it is restricted to particular dispersive models. A scheme for isotropic metasurfaces with Lorentzian dispersion was presented in~\cite{Shulabh_FDTD1}, but this scheme, based on the Auxiliary Differential Equation (ADE) approach~\cite{Susan_FDTD,Young_dispersive_FDTD}, requires solving a matrix equation at every march-on-time iteration. For more efficient and general analysis, an extension of the method of~\cite{GSTC_FDTD} to dispersive metasurfaces was developed in~\cite{Usef_dispersive}

This paper presents an overview of the research on the computational analysis of metasurfaces\footnote{This topic is a few years old at the time of this writing, and it is quite certain, considering the dramatically increasing interest in metasurfaces, that much more results will be reported in forthcoming years; so the reader is invited to follow up on developments subsequent to the publication of this overview.}. Section~\ref{sec:ms_as_sheet_discont} justifies the sheet discontinuity model for the treatment of metasurfaces. Section~\ref{sec:conv_BD_TC_issue} reviews conventional boundary and transition conditions and deduces their inapplicability to general metasurfaces, prompting for the need of a novel approach. This approach is the GSTC technique combined with a surface susceptibility tensor description of the metasurface medium, presented in Sec.~\ref{sec:KA_GSTC_Funds} as the foundation for metasurface model analysis. Section~\ref{sec:usefulness_GSTC_Analysis} explains the usefulness of the GSTC analysis in the holistic design of metasurfaces. Sections~\ref{sec:GSTC-FDFD} and~\ref{sec:GSTC-FDTD} subsequently overview the GSTC-based frequency-domain and time-domain computational techniques, respectively, reported to date. Finally, Sec.~\ref{sec:conclusion} draws conclusions and discusses prospective developments.

\section{Metasurface Bianisotropic Surface Susceptibility Model}\label{sec:ms_as_sheet_discont}

A readily available way to analyze a metasurface would be to handle it with an existing general-purpose full-wave electromagnetic simulator. However, since a metasurface is a highly complex structure -- being electrically thin, electrically large, composed of sub-wavelength particles with deeply subwavelength features, nonperiodic and bianisotropic -- as pointed out in Sec.~\ref{Sec:Intro} and illustrated in Fig.~\ref{Fig:Metasurface}, this brute-force approach would be essentially impractical: It would require huge memory resources and take prohibitive computation times while providing little insight into the physics of the metasurface. Therefore, an appropriate \emph{model} is essential for the efficient analysis of a metasurface.

What could be such a model? As mentioned in Sec.~\ref{Sec:Intro}, a metasurface has a deeply subwavelength thickness. Therefore, propagation or resonance effects in the direction perpendicular to its plane can be safely neglected (Fig.~\ref{Fig:Metasurface}). a metasurface is hence a \emph{local} entity: the field on its transmission side ($\boldsymbol{\psi}^\text{t}$) at the point $(x,y,0^+)$ depends on the field on its incidence-reflection side ($\boldsymbol{\psi}^\text{i}+\boldsymbol{\psi}^\text{r}$) only at the point $(x,y,0^-)$. Consequently, a metasurface may be safely modeled as a \emph{zero-thickness} sheet or sheet discontinuity. Moreover, as also mentioned in Sec.~\ref{Sec:Intro}, a metasurface is composed of a subwavelength particle lattice. It may be therefore locally homogenized, and hence described by a \emph{continuous} mathematical function, that is in general tensorial due to bianisotropy. Thus, a metasurface may be efficiently modeled as a surface (zero-thickness) continuous tensorial susceptibility function, $\te{\chi}(x,y)$, which is immensely simpler than the physical metasurface structure.

This model is extremely powerful, as it provides the foundation for:
\begin{itemize}
  \item a \emph{straightforward and analytical technique} for synthesis [$\raisebox{.5pt}{\textcircled{\raisebox{-.9pt} {1}}}$ in Fig.~\ref{Fig:Design}]~\cite{Karim_general_mts_syn,Guillaume_Refr_ms_no_diff};
  \item a \emph{fast and efficient scheme} for analysis [$\raisebox{.5pt}{\textcircled{\raisebox{-.9pt} {4}}}$ in Fig.~\ref{Fig:Design}]~\cite{GSTC_FDFD,GSTC_FDTD}, which is the focus of this paper;
  \item a \emph{unique insight into the physics} of metasurfaces [spatial and temporal variation, passive/active nature, reciprocity/nonreciprocity, tensorial structure (monoisotropy, monoanisotropy, bianisotropy), linearity/nonlinearity and transformation multiplicity]~\cite{Taravati_Nonrecip_nongyro_magless_ms, Eleftheriades_bianiso_ms,KA_2nd_order_mts}.
\end{itemize}

Of course, the $\te{\chi}(x,y)$ model will eventually have to be connected to the physical metasurface structure [$\raisebox{.5pt}{\textcircled{\raisebox{-.9pt} {2}}}$ and $\raisebox{.5pt}{\textcircled{\raisebox{-.9pt} {3}}}$ in Fig.~\ref{Fig:Design}], which is typically composed of 1-2 substrates and 2-3 metallization layers, with an overall subwavelength thickness~\cite{Laurin_Circ_Polar_3layer, Pfeiffer_Mts_3layer,Monticone_opti_Trans_contr_metascreen, Laurin_Trans_array_number_layers, limits_ultrathin_ms_2017}, supporting arrays of scattering particles with specific geometries. This connection may be accomplished by scattering parameter mapping: the physical/model parameters (particle geometries and lattice / susceptibility tensor structures and functions) are tuned to yield scattering parameters identical to those of the model/physical simulations for the synthesis~[$\raisebox{.5pt}{\textcircled{\raisebox{-.9pt} {2}}}$] / analysis~[$\raisebox{.5pt}{\textcircled{\raisebox{-.9pt} {3}}}$] operations. Using this technique, the metasurface is seen as a ``black box'',  where equivalence between the sheet model and the physical structure is automatically satisfied despite the fact that two entities have different thicknesses (zero thickness / thickness of the actual physical metasurface)~\cite{Karim_general_mts_syn}.

\section{Inapplicability of Conventional Boundary and Transition Conditions} \label{sec:conv_BD_TC_issue}

How could one possibly handle the surface susceptibility tensor metasurface model described in Sec.~\ref{sec:ms_as_sheet_discont}? Can one apply conventional Boundary Conditions (BCs) or Transition Conditions (TCs) to simulate such a complex discontinuity?

Conventional BCs are the usual boundary conditions at an interface between two media, including the Perfect Electric Conductor (PEC) BC, the Perfect Magnetic Conductor (PMC) BC, the Perfect Electromagnetic Conductor (PEMC) BC~\cite{Lindell_PEMC,Lindell_PEC,Caloz_PEMC}, the Absorbing Boundary Condition (ABC)~\cite{Volakis_FEM, MUR1981_FDTD_Abs_BD} and the Perfectly Matched Layers (PMLs) BC~\cite{BERENGER1994}. These BCs address the problem of a discontinuity formed by the juxtaposition of two different media. This does not correspond to the surface susceptibility tensor (zero-thickness) sheet model of a metasurface, which is a discontinuity per se, not requiring the presence of surrounding media, although such media can naturally be present. BCs are therefore inapplicable to the metasurface model since they cannot account for the presence of a sheet between media. This may be rigorously demonstrated as follows. The classical BCs at an interface between two media read
\begin{subequations}
\label{eq:CBC}
\begin{equation}
\ve{\hat{z}}\times\Delta\ve{{H}}
=\ve{{J}}_\text{s},\label{eq:CBC1}
\end{equation}
\begin{equation}
\Delta\ve{{E}}\times\ve{\hat{z}}
=\ve{{K}}_\text{s},\label{eq:CBC2}
\end{equation}
\begin{equation}
\ve{\hat{z}}\cdot\Delta\ve{{D}}
={\rho}_\text{s}^\textrm{e},\label{eq:CBC3}
\end{equation}
\begin{equation}
\ve{\hat{z}}\cdot\Delta\ve{{B}}
={\rho}_\text{s}^\textrm{m},\label{eq:CBC4}
\end{equation}
\end{subequations}
where $\ve{{J}}_\text{s}$, $\ve{{K}}_\text{s}$, ${\rho}_\text{s}^\textrm{e}$ and ${\rho}_\text{s}^\textrm{m}$ are the impressed electric current surface density, magnetic current surface density, electric charge surface density and magnetic charge surface density, respectively, while the operator $\Delta$ denotes the difference of the fields at both sides of the metasurface. Equations~\eqref{eq:CBC} are conventionally obtained from the integral form of Maxwell equations. The relations~\eqref{eq:CBC1} and~\eqref{eq:CBC2} are derived by applying Stokes theorem with integration over the path of a closed contour across the interface, while the relations~\eqref{eq:CBC3} and~\eqref{eq:CBC4} are derived by applying Gauss theorem with integration over the surface area of a pillbox at the interface between the two media. These BCs, found in most electromagnetic textbooks, cannot relate the fields across a metasurface sheet because they do not rigourously apply to an interface supporting currents and charges, as pointed out by Schelkunoff about 50 years ago~\cite{Schelkunoff_EM_theory}. Consider for example the discontinuity in the displacement vector, $\ve{{D}}$, in~\eqref{eq:CBC3}. This equation is stricto senso incorrect, for the following two reasons: first, Gauss theorem is applicable \emph{only} if $\ve{{D}}$ is continuous within the volume of integration, which is obviously not the case when $\rho_\text{s}^\textrm{e}\neq 0$. The relation is correct everywhere except at the interface (up to $z=0^\pm$), but fails to describe the behavior of the field at $z=0$. Second, Eq.~\eqref{eq:CBC3}, implying that $\ve{{D}}$ is perfectly continuous in the absence of impressed electric surface charges, fails to consider the contribution of excitable dipole or higher-order multipole moments or polarization currents, that typically model a metasurface. These issues also apply to the other three relations in~\eqref{eq:CBC}. Conventional boundary conditions are thus not capable to properly account for the presence of a metasurface.

In contrast to BCs, TCs, which are fictitious sheet discontinuities used in the computation of thin structures such as coating films~\cite{Sergey_Modeling_Appl_EM}, frequency selective surfaces~\cite{RajMittra_FSS_Analysis} and two-dimensional materials (graphene, black phosphorous, etc.)~\cite{Nima_graphene_coupler, Nayyeri_FDTD, Nima_THtz_Graphene_PN_junc}, \emph{do} model zero-thickness discontinuities and would therefore a priori seem applicable to the metasurface model. However, conventional TCs can only account for discontinuity in the electric field \emph{or} in the magnetic field, and \emph{not both}. Indeed, they have the form impedance boundary condition (IBC) form
\begin{subequations}\label{eq:IBCs}
\begin{equation}
\Delta\ve{E} = \bar{\bar{\mathbf{Z}}} \cdot \ve{H},
\end{equation}
or
\begin{equation}
\Delta\ve{H} = \bar{\bar{\mathbf{Y}}} \cdot \ve{E},
\end{equation}
\end{subequations}
where $\Delta$ represents the field discontinuity across the sheet and $\bar{\bar{\mathbf{Z}}}$ and $\bar{\bar{\mathbf{Y}}}$ are in general 4$\times$4 immittance tensors, corresponding to electric only and magnetic only discontinuity, respectively. However, metasurfaces generally represent simultaneously electric and magnetic discontinuities. Moreover, they are generally bianisotropic, which is incompatible with the aforementioned mathematical forms.

Thus, neither BCs nor conventional TCs can deal with the metasurface sheet susceptibility model, and another type of TC is therefore required. Fortunately, such a TC, providing exactly what is needed, exists: It is the Generalized Sheet Transition Conditions (GSTCs), that will be described in the next section.
\section{GSTC Fundamentals}\label{sec:KA_GSTC_Funds}
The GSTCs were initially developed by Idemen~\cite{Idemen} and later applied to metasurfaces by Kuester {\it et al.}~\cite{Kuester}. They are established by expressing the discontinuities of the fields at the sheet in terms of an expansion over derivatives of the Dirac delta distribution~\cite{Distribution_theory}\footnote{It is also possible to derive the GSTCs using a more conventional technique, based on pillbox integration~\cite{Albooyeh_bianis_Mts_EM_charac}.}. Such an expansion generally takes the form
\begin{equation}
\label{eq:General_F}
f(z)=\left \{ f(z) \right\}+\sum_{k=0}^{N}f_{k}\delta^{(k)}(z),
\end{equation}
where $f(z)$ corresponds to any quantity in Maxwell equations, $\delta^{(k)}$ represents the $k$-th derivative of the Dirac delta distribution, $\left \{ f(z) \right\}$ represents the \emph{regular part} of the function $f(z)$, which corresponds to $f(z)$ everywhere in space except at $z=0$, and the sum term in~\eqref{eq:General_F} corresponds to the \emph{singular part} of $f(z)$, which represents the value of $f(z)$ precisely at $z=0$.

Substituting all the quantities in Maxwell equations with their expression in the form of~\eqref{eq:General_F} leads to two sets of equations: the \emph{universal boundary conditions} and the \emph{compatibility relations}~\cite{Idemen, Karim_general_mts_syn}. Recursively solving the latter for each value of $k$ in~\eqref{eq:General_F} leads to the final expression of the GSTCs, which read, in the time-harmonic regime,
\begin{subequations}
\label{eq:GSTC}
\begin{equation}
\ve{\hat{z}}\times\Delta\ve{H}
=\ve{J}_\text{s} + \frac{\partial\ve{P}_{\parallel}}{\partial t}-\ve{\hat{z}}\times\nabla_{\parallel}M_{z},
\end{equation}
\begin{equation}
\Delta\ve{E}\times\ve{\hat{z}}
=\ve{K}_\text{s}+j\omega\mu_0 \frac{\partial\ve{M}_{\parallel}}{\partial t}-\nabla_{\parallel}\bigg(\frac{P_{z}}{\epsilon_0 }\bigg)\times\ve{\hat{z}},
\end{equation}
\begin{equation}
\ve{\hat{z}}\cdot\Delta\ve{D}
=\rho_\text{s}^\text{e}-\nabla\cdot\ve{P}_{\parallel},
\end{equation}
\begin{equation}
\ve{\hat{z}}\cdot\Delta\ve{B}
=\rho_\text{s}^\text{m}-\mu_0 \nabla\cdot\ve{M}_{\parallel}.
\end{equation}
\end{subequations}
In these relations, the spectral versions of the polarization vector densities, $\ve{\tilde{P}}$ and $\ve{\tilde{M}}$, are related to the total fields via the surface susceptibility tensor spectral functions  $\te{\tilde{\chi}}_\text{ee}$, $\te{\tilde{\chi}}_\text{mm}$, $\te{\tilde{\chi}}_\text{em}$ and $\te{\tilde{\chi}}_\text{me}$\footnote{The \emph{surface} susceptibilities are expressed in meters, as may be seen by comparing~\eqref{eq:GSTC} and~\eqref{eq:PM}.} by the constitutive relations
\begin{subequations}
\label{eq:PM}
\begin{equation}
 \ve{\tilde{P}}=\epsilon_0 \te{\tilde{\chi}}_\text{ee}\cdot\ve{\tilde{E}}+\frac{1}{c_0}\te{\tilde{\chi}}_\text{em}\cdot\ve{\tilde{H}},
\end{equation}
\begin{equation}
 \ve{\tilde{M}}=\te{\tilde{\chi}}_\text{mm}\cdot\ve{\tilde{H}}+\frac{1}{\eta_0}\te{\tilde{\chi}}_\text{me}\cdot\ve{\tilde{E}}
 \footnote{In these last relations, the electric and magnetic fields are rigorously the fields \emph{acting} at the position of the metasurface. However, these fields can be approximated as the arithmetic average of the fields before and after the metasurface, i.e. $\tilde{\boldsymbol{\psi}}_\text{act}\approx\tilde{\boldsymbol{\psi}}_\text{av} = (\tilde{\boldsymbol{\psi}}_\text{i}+\tilde{\boldsymbol{\psi}}_\text{r}+\tilde{\boldsymbol{\psi}}_\text{t})/2=\tilde{\boldsymbol{\psi}}$.},
\end{equation}
\end{subequations}
where the tilde symbol indicates spectral quantities.

The GSTCs used in metasurface so far, and throughout this paper, are those corresponding to reduction of the sum in~\eqref{eq:General_F} to the single term $N=0$ ($f_k=0$ for $k>0$). In that case, the boundary conditions account for the discontinuity of the fields but not for discontinuities in their derivatives\footnote{This is sufficient for most metasurfaces. However, there are cases where such restriction would create problems. For instance, a metasurface transforming the incident field into a transmitted field corresponding to the phase-reversed version of the incident field could not be described by a series truncated to $N=0$ in~\eqref{eq:General_F}. Indeed, the corresponding quantities would include only the even $\delta(z)$ distribution whereas the field transformation is obviously odd in nature. In such a case, one should at least include the term $N=1$ to involve the odd distribution $\delta'(z)$. The application of higher-order GSTCs to metasurfaces is still an open research topic.}.

It is instructive to compare the GSTCs [Eqs.~\eqref{eq:GSTC}] with the conventional BCs [Eqs.~\eqref{eq:CBC}] and IBCs [Eqs.~\eqref{eq:IBCs}]. The first comparison show that GSTCs are an extended version of the conventional BCs, where the metasurface sheet discontinuity is accounted for by added polarization currents, associated with the bianisotropic surface susceptibilities that characterize it~[Eqs.~\eqref{eq:PM}]. The second comparison show that the GSTCs, in addition to being more rigorous, are also more complete than the IBCs, since they take into account possible normal susceptibility components while the IBCs only consider tangential impedance/admittance components and ignores normal polarization\footnote{For instance, IBCs cannot directly account for the magnetic polarization ($\tilde{M}_z \neq 0$) caused by planar metallic rings, whereas the GSTCs can.}. Moreover, contrary to the IBCs, the GSTCs can be straightforwardly extended to include nonlinear susceptibility terms~\cite{KA_2nd_order_mts}, as well as possibly higher-order discontinuity terms.

In the particular case, covering a large number of practical metasurfaces, where $\tilde{M}_z=\tilde{P}_z=0$\footnote{The case $\tilde{M}_z,\tilde{P}_z\neq 0$ leads to differential equations and is therefore more problematic~\cite{Karim_general_mts_syn}.} and hence the susceptibility tensors reduce to $2\times 2$ transverse tensors, one obtains, by substituting~\eqref{eq:PM} into the spectral version of~\eqref{eq:GSTC} (monochromatic regime), the particularly convenient closed-form relations
\begin{subequations}\label{eq:Freq_GSTC}
  \begin{align}\label{eq:Freq_GSTC1}
  \left(
  \begin{array}{c}
    -\Delta \tilde{H}_y \\
    \Delta \tilde{H}_x \\
  \end{array}
\right) &=j\omega\varepsilon_0 \left(
                        \begin{array}{cc}
                          \tilde{\chi} _{\textrm{ee}}^{xx} & \tilde{\chi} _{\textrm{ee}}^{xy} \\
                          \tilde{\chi} _{\textrm{ee}}^{yx} & \tilde{\chi} _{\textrm{ee}}^{yy} \\
                        \end{array}
                      \right)
                      \left(
                        \begin{array}{c}
                          \tilde{E}_{x,\textrm{av}} \\
                          \tilde{E}_{y,\textrm{av}} \\
                        \end{array}
                      \right)\\\notag
                      &+j\omega\sqrt{\varepsilon_0 \mu_0} \left(
                                                           \begin{array}{cc}
                                                             \tilde{\chi} _{\textrm{em}}^{xx} & \tilde{\chi} _{\textrm{em}}^{xy} \\
                                                             \tilde{\chi} _{\textrm{em}}^{yx} & \tilde{\chi} _{\textrm{em}}^{yy} \\
                                                           \end{array}
                                                         \right)
                                                         \left(
                                                           \begin{array}{c}
                                                             \tilde{H}_{x,\textrm{av}} \\
                                                             \tilde{H}_{y,\textrm{av}} \\
                                                           \end{array}
                                                         \right),
  \end{align}
  \begin{align}\label{eq:Freq_GSTC2}
  \left(
  \begin{array}{c}
    \Delta \tilde{E}_y \\
    -\Delta \tilde{E}_x \\
  \end{array}
\right) &=j\omega\mu_0 \left(
                        \begin{array}{cc}
                          \tilde{\chi} _{\textrm{mm}}^{xx} & \tilde{\chi} _{\textrm{mm}}^{xy} \\
                          \tilde{\chi} _{\textrm{mm}}^{yx} & \tilde{\chi} _{\textrm{mm}}^{yy} \\
                        \end{array}
                      \right)
                      \left(
                        \begin{array}{c}
                          \tilde{H}_{x,\textrm{av}} \\
                          \tilde{H}_{y,\textrm{av}} \\
                        \end{array}
                      \right)\\\notag
                      &+j\omega\sqrt{\varepsilon_0 \mu_0} \left(
                                                           \begin{array}{cc}
                                                             \tilde{\chi}_{\textrm{me}}^{xx} & \tilde{\chi} _{\textrm{me}}^{xy} \\
                                                             \tilde{\chi}_{\textrm{me}}^{yx} & \tilde{\chi} _{\textrm{me}}^{yy} \\
                                                           \end{array}
                                                         \right)
                                                         \left(
                                                           \begin{array}{c}
                                                             \tilde{E}_{x,\textrm{av}} \\
                                                             \tilde{E}_{y,\textrm{av}} \\
                                                           \end{array}
                                                         \right).
\end{align}
\end{subequations}

\section{Usefulness of the Model-Based Analysis}\label{sec:usefulness_GSTC_Analysis}

As mentioned in Sec.~\ref{Sec:Intro}, this paper does not deal with the complete analysis operation, represented by the large vertical up arrow in Fig.~\ref{Fig:Design}, but essentially with the model-to-field part of the analysis [$\raisebox{.5pt}{\textcircled{\raisebox{-.9pt} {4}}}$], leaving out the physical-structure-to-model analysis [$\raisebox{.5pt}{\textcircled{\raisebox{-.9pt} {3}}}$ in Fig.~\ref{Fig:Design}]. The reason for this is twofold. First, the only existing approach for $\raisebox{.5pt}{\textcircled{\raisebox{-.9pt} {3}}}$ (as $\raisebox{.5pt}{\textcircled{\raisebox{-.9pt} {4}}}$) is the scattering parameter mapping technique discussed in the last paragraph of Sec.~\ref{sec:ms_as_sheet_discont}, which is tedious but straightforward. Second, $\raisebox{.5pt}{\textcircled{\raisebox{-.9pt} {4}}}$ is tremendously useful in the holistic design shown in Fig.~\ref{Fig:Design}.

\begin{figure}[!ht]
\centering
\begin{subfigure}{0.5\columnwidth}
  \centering
  \includegraphics[width=1\columnwidth]{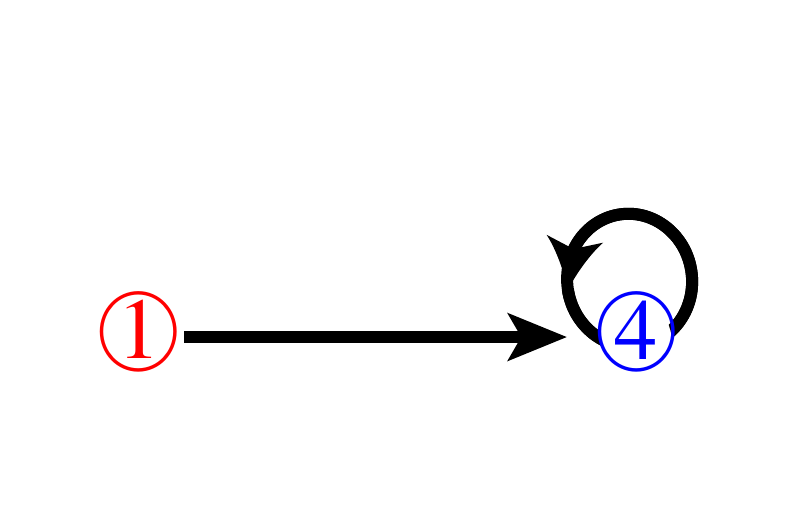}
  \caption{} \label{Fig:Model_based_useful_A}
\end{subfigure}%
\begin{subfigure}{0.5\columnwidth}
\centering
\includegraphics[width=1\columnwidth]{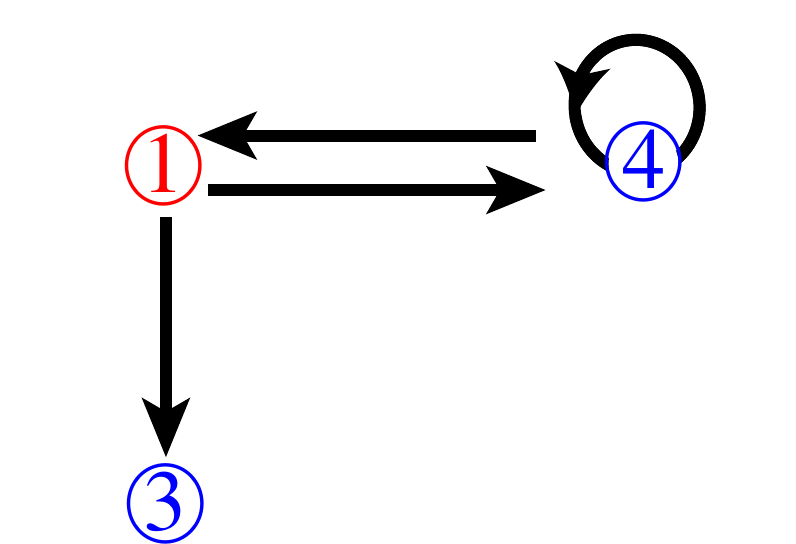}
  \caption{} \label{Fig:Model_based_useful_B}
\end{subfigure}

\begin{subfigure}{0.5\columnwidth}
  \centering
  \includegraphics[width=1\columnwidth]{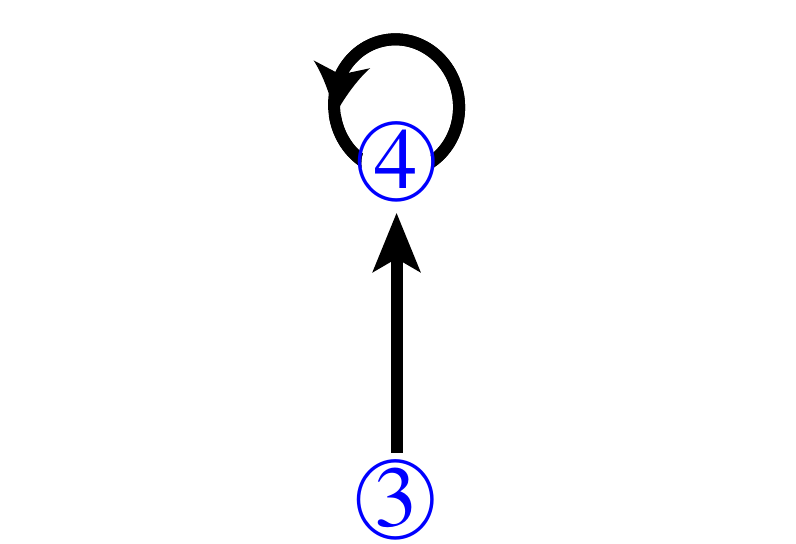}
  \caption{} \label{Fig:Model_based_useful_C}
\end{subfigure}%
\begin{subfigure}{0.5\columnwidth}
\centering
\includegraphics[width=1\columnwidth]{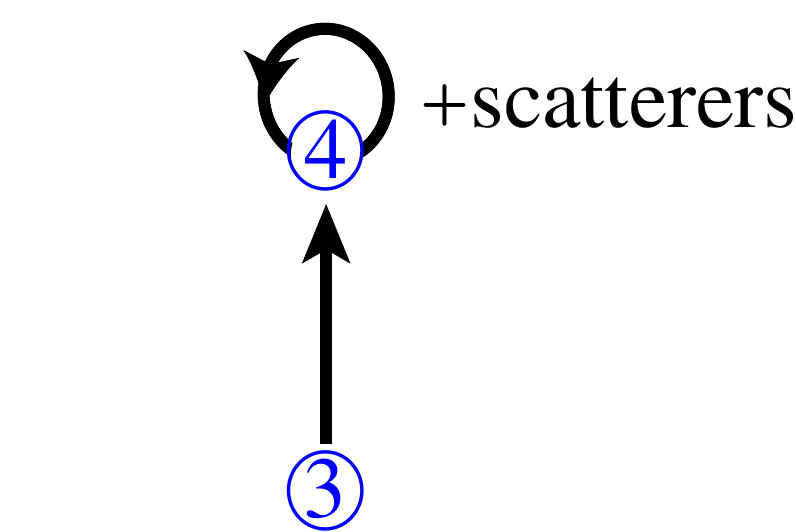}
  \caption{} \label{Fig:Model_based_useful_D}
\end{subfigure}
\caption{Usefulness of the model-based analysis [$\textcircled{4}$ in Fig.~\ref{Fig:Design}]. (a) Synthesis verification and characterization. (b) Synthesis by synthesis-analysis iterations. (c) Analysis and characterization from extracted susceptibilities. (d) Same as (c) but in the presence of scatterers.}
\label{Fig:Model_based_useful}
\end{figure}

Figure~\ref{Fig:Model_based_useful} represents four operations where the model based analysis [$\raisebox{.5pt}{\textcircled{\raisebox{-.9pt} {4}}}$] is useful in the design of a metasurface:
\begin{itemize}
  \item Fig~\ref{Fig:Model_based_useful_A}: Once a metasurface has been synthesized in terms of its susceptibility model [$\raisebox{.5pt}{\textcircled{\raisebox{-.9pt} {1}}}$], $\raisebox{.5pt}{\textcircled{\raisebox{-.9pt} {4}}}$ allows one to \emph{verify} that it produces the specified fields and to \emph{characterize} it for parameters (frequency, angle, polarization and waveform of the incident wave; metasurface size) other than the specified ones. The latter is essential to understand how the metasurface will behave and may affect other objects in a practical situation.
  \item Fig~\ref{Fig:Model_based_useful_B}: $\raisebox{.5pt}{\textcircled{\raisebox{-.9pt} {4}}}$ allows efficient synthesis under the form of iterative synthesis-analysis. For instance, if the result of $\raisebox{.5pt}{\textcircled{\raisebox{-.9pt} {1}}}$ is found inappropriate (e.g. undesired lossy/active or nonreciprocal susceptibilities, unpractically fast variations in terms of wavelength, etc.), $\raisebox{.5pt}{\textcircled{\raisebox{-.9pt} {4}}}$ may help in adjusting the design, either by transforming tensorial structure of the metasurface (e.g. from monoisotropic to bianisotropic~\cite{Guillaume_Refr_ms_no_diff}, etc.) or by relaxing specifications (e.g. allowing tolerable loss, increasing the size, etc.). Once this iterative operation has been performed to satisfaction, one moves on to the physical structure design [$\raisebox{.5pt}{\textcircled{\raisebox{-.9pt} {3}}}$].
  \item Fig~\ref{Fig:Model_based_useful_C}: After the susceptibilities of a physical metasurface structure have been extracted, for a minimal set of test waves necessary to determine all its tensorial susceptibility parameters, $\raisebox{.5pt}{\textcircled{\raisebox{-.9pt} {4}}}$ allows to efficiently characterize the metasurface for \emph{any other} parameters than the specified ones, without having to simulate the complex physical structure.
  \item Fig~\ref{Fig:Model_based_useful_D}: The previous operation is all the more beneficial when \emph{scatterering objects} co-exist with the metasurface, as the computational burden will then be essentially restricted to these objects rather than including, possibly prohibitively, the metasurface physical structure.
\end{itemize}

Now that the usefulness of the analysis based on the bianisotropic surface susceptibility model has been explained and that the GSTCs required to handle this model have been described (Sec.~\ref{sec:KA_GSTC_Funds}), we may overview the computational techniques developed to simulate metasurfaces at that level [$\raisebox{.5pt}{\textcircled{\raisebox{-.9pt} {4}}}$ in Fig.~\ref{Fig:Design}]. These techniques essentially consist in grafting GSTC-based schemes into conventional numerical algorithms, both in the frequency domain and in the time domain. Therefore, the fundamental features of the corresponding frequency-domain and time-domain methods are maintained~\cite{Garg,Bondeson}, and we shall delve here only into metasurface-specific aspects of the computational techniques overviewed.

\section{GSTC-Based Frequency-Domain Computational Techniques}\label{sec:GSTC-FDFD}

We shall here describe the implementation of the GSTCs first in the Finite-Difference Frequency-Domain (FDFD) method, in some details and with an illustrative example, and next, briefly, in the Spectral-Domain Integral-Equation (SD-IE) method and in the Finite Element Method (FEM).

\subsection{Finite Difference Frequency Domain (FDFD)}

For simplicity, we consider the TM$_z$ 2D ($\partial/\partial y=0$) problem, whose only nonzero field components are $\tilde{H}_y$, $\tilde{E}_x$ and $\tilde{E}_z$. An essentially similar procedure may be applied to the TE$_z$ and 3D problems. In the 2D TM$_z$ case, Eqs.~\eqref{eq:Freq_GSTC} reduce to
\begin{subequations}\label{TE_GSTC}
\begin{equation}\label{TE_GSTC1}
    -\Delta \tilde{H}_y=j\omega\varepsilon_0\tilde{\chi}_\textrm{ee}^{xx}\tilde{E}_{x,\textrm{av}}+jk_0\tilde{\chi}_\textrm{em}^{xy}\tilde{H}_{y,\textrm{av}},
\end{equation}
\begin{equation}\label{TE_GSTC2}
    -\Delta  \tilde{E}_x=j\omega\mu_0\tilde{\chi}_\textrm{mm}^{yy}\tilde{H}_{y,\textrm{av}}+jk_0\tilde{\chi}_\textrm{me}^{yx}\tilde{E}_{x,\textrm{av}}.
\end{equation}
\end{subequations}

The key question is how to position the metasurface in the FDFD computational grid. Positioning it on E-field or H-field nodes would allow only magnetic-field discontinuity or electric-field discontinuity~\cite{Nayyeri_FDTD}, respectively, and not both, since only one field quantity is specified at a given node, which prevents inserting any discontinuity in this field.

\begin{figure}[!ht]
\centering
\includegraphics[width=1\columnwidth]{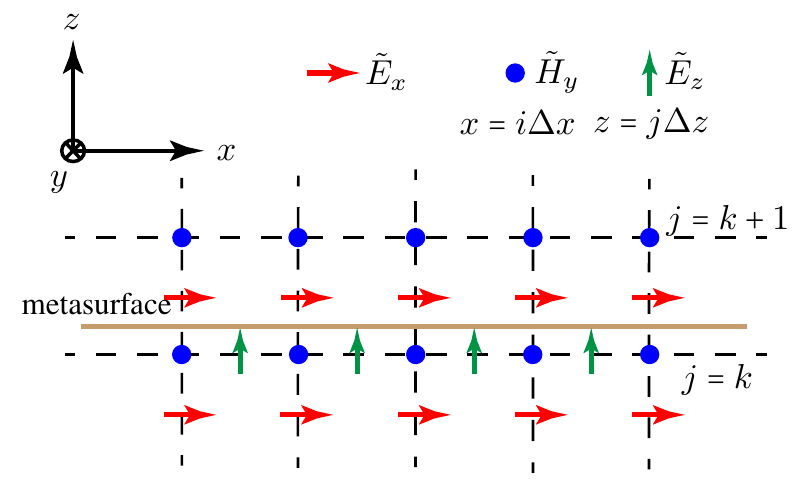}
  \caption{Metasurface positioning in the FDFD grid at $z=(k+\frac{1}{4})\Delta z$ along the $z$-direction.}
  \label{Fig:FDFD_grid}
\end{figure}

A solution, proposed in~\cite{GSTC_FDFD_APS,GSTC_FDFD}, consists in positioning the metasurface \emph{between} adjacent nodes, as shown in Fig.~\ref{Fig:FDFD_grid}, and to apply the GSTCs to that position, while using the conventional FDFD scheme everywhere else.

The standard FDFD equations for TM$_z$ are~\cite{Rumpf_FDFD2012}
\begin{subequations}
\begin{align}\label{eq:FDFD_Ey}
  -\frac{\tilde{H}_y^{i,j+1}-\tilde{H}_y^{i,j}}{\Delta z}= j\omega\varepsilon_0\tilde{\varepsilon}_{xx}^{i,j+\frac{1}{2}}\tilde{E}_x^{i,j+\frac{1}{2}},
\end{align}
\begin{align}\label{eq:FDFD_Hy}
   \frac{\tilde{E}_x^{i,j+\frac{1}{2}}-\tilde{E}_x^{i,j-\frac{1}{2}}}{\Delta z}-&\frac{\tilde{E}_z^{i+\frac{1}{2},j}-\tilde{E}_z^{i-\frac{1}{2},j}}{\Delta x}\\\notag &=j\omega\mu_0\tilde{\mu}_{yy}^{i,j}\tilde{H}_y^{i,j},
\end{align}
\begin{align}\label{eq:FDFD_Ez}
   \frac{\tilde{H}_y^{i+1,j}-\tilde{H}_y^{i,j}}{\Delta x}= j\omega\varepsilon_0\tilde{\varepsilon}_{zz}^{i+\frac{1}{2},j}\tilde{E}_z^{i+\frac{1}{2},j},
\end{align}
\end{subequations}
which may be recast, for the entire computational domain, into the compact matrix system
\begin{subequations}\label{eq:FDFD_matrix}
\begin{equation}\label{eq:FDFD_Ex_matrix}
  -\ve{\tilde{D}}_\textrm{h}^z\ve{\tilde{H}}_y=\ve{\tilde{\varepsilon}}_{xx}\ve{\tilde{E}}_x,
\end{equation}
\begin{equation}\label{eq:FDFD_Hy_matrix}
  \ve{\tilde{D}}_\textrm{e}^z\ve{\tilde{E}}_x-\ve{\tilde{D}}_\textrm{e}^x\ve{\tilde{E}}_z=\ve{\tilde{\mu}}_{yy}\ve{\tilde{H}}_y,
\end{equation}
\begin{equation}\label{eq:FDFD_Ez_matrix}
  \ve{\tilde{D}}_\textrm{h}^x\ve{\tilde{H}}_y=\ve{\tilde{\varepsilon}}_{zz}\ve{\tilde{E}}_z,
\end{equation}
\end{subequations}
where $\ve{\tilde{D}}_\textrm{h}^z, \ve{\tilde{D}}_\textrm{e}^z, \ve{\tilde{D}}_\textrm{e}^x$ and $\ve{\tilde{D}}_\textrm{h}^x$ are differential matrix operators, $\ve{\tilde{\varepsilon}}_{xx}, \ve{\tilde{\varepsilon}}_{zz}$ and $\ve{\tilde{\mu}}_{yy}$ are permittivity and permeability matrices, respectively, and $\ve{\tilde{E}}$ and $\ve{\tilde{H}}$ are field column-vectors, respectively\footnote{Details on the resolution of~\eqref{eq:FDFD_matrix} are provided in~\cite{Rumpf_FDFD2012}. Essentially, the matrix system, upon elimination of $\ve{\tilde{E}}_x$ and $\ve{\tilde{E}}_z$, is transformed into a matrix equation of the form $\ve{\tilde{A}}\ve{\tilde{H}}_y=\ve{\tilde{S}}$, where $\ve{\tilde{A}}=\ve{\tilde{D}}_e^z\ve{\tilde{\varepsilon}}_{xx}^{-1}\ve{\tilde{D}}_h^z+\ve{\tilde{D}}_e^x\ve{\tilde{\varepsilon}}_{zz}^{-1}\ve{\tilde{D}}_h^x-\ve{\tilde{\mu}}_{yy}$ is the coefficient matrix and $\ve{\tilde{H}}_y$ and $\ve{\tilde{S}}$ are the unknown field vector and the source vector [added to~\eqref{eq:FDFD_matrix}], respectively.\label{fn:FDTD_matrix_syst}}~\cite{Rumpf_FDFD2012}.

\begin{figure}[!ht]
\centering
\begin{subfigure}{1\columnwidth}
  \centering
  \includegraphics[trim=4mm 0 0 0, clip=true,width=1\columnwidth]{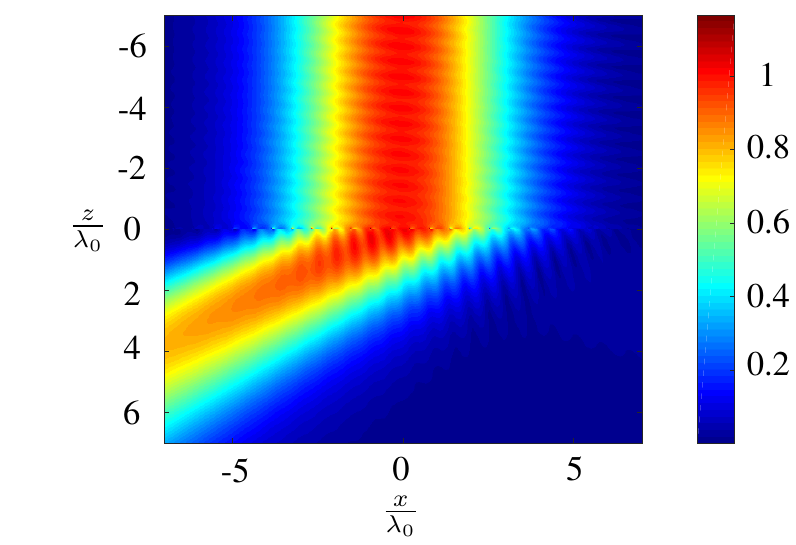}
  \caption{} \label{Fig:FDFD_ABS}
\end{subfigure}

\begin{subfigure}{1\columnwidth}
  \centering
  \includegraphics[width=1\columnwidth]{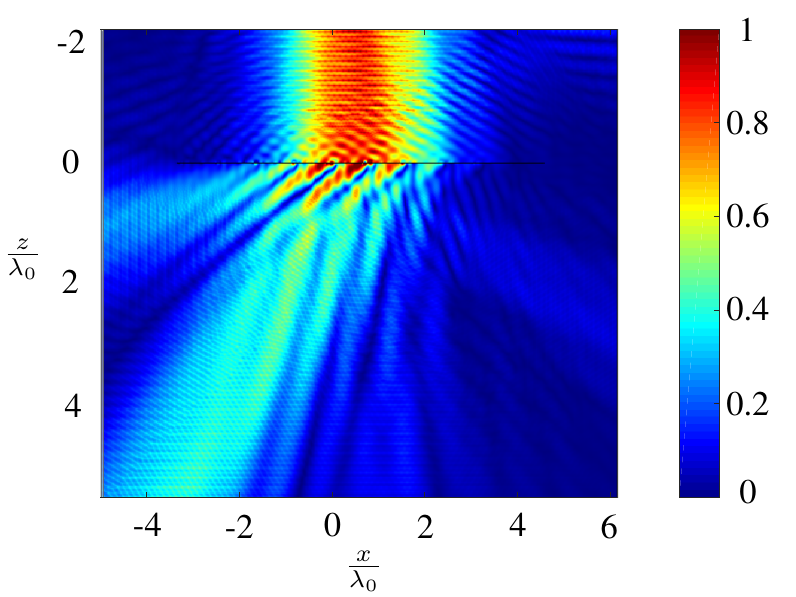}
  \caption{} \label{Fig:COMSOL_ABS}
\end{subfigure}
\caption{Illustrative result of the GSTC-FDFD scheme for the problem of a reflection-less monoisotropic metasurface, placed at $z=0$, refracting a modulated Gaussian beam incident at $\theta^\textrm{i}=0^\circ$ to the direction $\theta^\textrm{t}=60^\circ$. (a)~GSTC-FDFD simulation, exactly matching the synthesis specification [$\textcircled{1}$ in Fig.~\ref{Fig:Design}]. (b)~COMSOL simulation with diluted volume susceptibility in a thin slab model, with spurious diffraction orders due to slab resonances.}
\label{Fig:GSTCFDFD}
\end{figure}

For $j=k$ in~\eqref{eq:FDFD_Ey} $\tilde{E}_x^{i,k+\frac{1}{2}}$ involves the fields $\tilde{H}_y^{i,k+1}$ and $\tilde{H}_y^{i,k}$, positioned on either side of the metasurface (Fig.~\ref{Fig:FDFD_grid}), and would therefore fail to account for the presence of the metasurface, since it considers only quantities located at $z=k\Delta z$ and $z=(k+1)\Delta z$ and ignores anything that could exist in between. To properly account for the effect of the metasurface, one must relate the fields on either side of it by the GSTCs. This may be accomplished by replacing~\eqref{eq:FDFD_Ey} with an approximate\footnote{This equation is only an \emph{approximation} of ~\eqref{TE_GSTC1}, since the involves fields are sampled at \emph{four different} positions, imposed by the FDFD staggered grid structure (four vertical positions [two for $\tilde{E}_x$ and two for $\tilde{H}_y$] in Fig.~\ref{Fig:FDFD_grid}), whereas Eq.~\eqref{TE_GSTC1} assumes zero thickness. In fact, the scheme, as detailed in~\cite{GSTC_FDFD}, spatially merges the fields $\tilde{E}_x^{i,k+1/2}$ and $\tilde{H}_y^{i,k}$, but the total distance between the two other sampled fields ($\tilde{H}_y^{i,k+1}$ and $\tilde{E}_x^{i,k-1/2}$) is one computational cell ($\Delta z$), which is still nonzero. However, this approximation does \emph{not} compromise the efficiency of the method. Indeed, only the area surrounding the metasurface sheet, between the $\tilde{H}_y^{i,k+1}$ and $\tilde{E}_x^{i,k-1/2}$ rows, is meshed, so that the overall computational mesh only depends on the surrounding media. In contrast, an equivalent conventional thin slab (as used in the COMSOL simulation in Fig.~\ref{Fig:COMSOL_ABS}) needs to be meshed and therefore involves, given the deeply subwavelength nature of the metasurface ($\delta\sim\lambda/100$), extremely dense meshing in the vicinity of the slab, which requires large memory resources and long computation time.\label{fn:approx}} discretized version of~\eqref{TE_GSTC1}\footnote{Using~\eqref{TE_GSTC2} would also be possible, but this leads to a formulation that is numerically less accurate.}, which yields, after grouping similar field terms,
\begin{equation}\label{TE_GSTC2_discret}
  \tilde{\alpha}_\textrm{m}^-\tilde{H}_y^{i,k+1}+\tilde{\alpha}_\textrm{m}^+\tilde{H}_y^{i,k}
  =\frac{j\omega\varepsilon_0\tilde{\chi}_\textrm{ee}^{xx}}{2}\left[\tilde{E}_x^{i,k+\frac{1}{2}}+\tilde{E}_x^{i,k-\frac{1}{2}} \right],
\end{equation}
where $\tilde{\alpha}_\textrm{m}^\pm=\pm1-\frac{jk_0\tilde{\chi}_\textrm{em}^{xy}}{2}$. Equation~\eqref{TE_GSTC2_discret} is to replace~\eqref{eq:FDFD_Ey} at the position $(i\Delta x,(k+\frac{1}{2})\Delta z)$ ($\forall i$). This substitution only changes some entries of $\ve{\tilde{D}}_\textrm{h}^z$ and $\ve{\tilde{\varepsilon}}_{xx}$ in~\eqref{eq:FDFD_Ex_matrix}~\cite{GSTC_FDFD}, without altering the form of this matrix equation.

Similarly, Eq.~\eqref{TE_GSTC2} may be discretized and rearranged to yield
\begin{equation}\label{TE_GSTC1_discret}
  \tilde{\alpha}_\textrm{e}^-\tilde{E}_x^{i,k+\frac{1}{2}}+\tilde{\alpha}_\textrm{e}^+\tilde{E}_x^{i,k-\frac{1}{2}}
  =\frac{j\omega\mu_0\tilde{\chi}_\textrm{mm}^{yy}}{2}\left[\tilde{H}_y^{i,k}+\tilde{H}_y^{i,k+1} \right],
\end{equation}
where $\tilde{\alpha}_\textrm{e}^\pm=\pm1-\frac{jk_0\tilde{\chi}_\textrm{me}^{zy}}{2}$, which is to replace~\eqref{eq:FDFD_Hy} at the position $(i\Delta x, k\Delta z)$ ($\forall i$). This only changes some entries of $\ve{\tilde{D}}_\textrm{e}^z, \ve{\tilde{D}}_\textrm{e}^x$ and $\ve{\tilde{\mu}}_{yy}$ in~\eqref{eq:FDFD_Hy_matrix}, as detailed in~\cite{GSTC_FDFD}, without altering the form of this matrix equation.

Equation~\eqref{eq:FDFD_Ez} is not affected by the presence of the metasurface, since its sampled fields are all located at the same side of it. Consequently,
Eq.~\eqref{eq:FDFD_Ez_matrix} is also unaffected.

Figure~\ref{Fig:GSTCFDFD} provides and illustrative result for the GSTC-FDFD scheme compared to that obtained using a thin-slab model for a refractive metasurface. The GSTC-FDTD result, shown in Fig.~\ref{Fig:FDFD_ABS}, is in close (although not perfect, due to the approximation mentioned in Footnote~\ref{fn:approx}) agreement with the specification, which demonstrates the validity of the proposed approach. The COMSOL thin-slab model consists in diluting the surface susceptibility in a deeply subwavelength ($\delta=\lambda/100$) slab into a volume susceptibility ($\chi_\text{vol}=\chi_\text{surf}/\delta$). The corresponding result, plotted in Fig.~\ref{Fig:COMSOL_ABS}, in addition to being computationally expensive, fails to reproduce the specified fields. This failure is attributed to the incapability of the thin-film model to account for rapid susceptibility variations (that turn out to be here very sharp on the wavelength scale~\cite{Guillaume_Refr_ms_no_diff}).

\subsection{Spectral-Domain Integral-Equation (SD-IE)}
GSTCs can also be implemented in the form of integral equations. In the case of a flat metasurface\footnote{Not all the metasurfaces reported to date are flat, but the vast majority of them are.}, the GSTC integral equations can be particularly conveniently manipulated in the spectral domain, where they do not require any field expansion and testing, contrary to the spatial case, and hence provide a much simpler formulation~\cite{NC_SD_IE}.

The corresponding Spectral-Domain Integral-Equation (SD-IE) method consists in representing all the fields and susceptibilities in terms of their frequency-domain ($\omega$) and transverse spectral-domain ($\mathbf{k}_T=k_x\mathbf{\hat{x}}+k_y\mathbf{\hat{y}}$) counterparts, whence the derivatives reduce to products, and substituting these expressions into the GSTCs, so as to obtain a system of coupled algebraic equations, whose solution provides the scattered fields~\cite{NC_SD_IE}.

The main advantage of this approach is that it reduces the original metasurface problem to a problem that is one dimension smaller (2D from 3D or 1D from 2D), since the need for meshing in the direction normal to the metasurface~\footnote{Incidentally, the absence of discretization in that direction results, in contrast to the FDFD case, in a computation that exactly models the zero-thickness of the metasurface in~\ref{eq:Freq_GSTC}.} is obviated by the angular-spectrum representation of the fields. This dramatically reduces the computational complexity and memory requirements. However, this is at the cost of restricting the range of tractable problems to those without additional scattering objects, unless these objects are transversally uniform, such as flat multilayer structures.

\subsection{Finite Element Method (FEM)}

GSTCs may also be implemented in the Finite Element Method (FEM), as reported in~\cite{GSTC_FEM}. This implementation essentially consists in linking the Neumann boundary conditions belonging to the domains on either side of the metasurface with the GSTC equations and surface susceptibility tensors\footnote{This linkage also results in a computation that exactly models the zero-thickness of the metasurface in~\ref{eq:Freq_GSTC}.}. Details may be found in~\cite{GSTC_FEM}, which treats both cases of a 1D `metasurface' and a 2D bianisotropic metasurface using triangular elements and linear basis functions.

\section{GSTC-Based Time-Domain \\ Computational Techniques}\label{sec:GSTC-FDTD}

We shall now describe the implementation of GSTCs in the Finite-Difference Time-Domain (FDTD) method for general space-varying and time-varying metasurfaces. We will first introduce the concept of virtual nodes, derive and interpret the subsequent key equations, and present an illustrative example. We will next explain how to take into account specific metasurface frequency dispersions.

\subsection{Finite Difference Time Domain (FDTD) Algorithm}\label{sec:FDTD}

In order to provide a global presentation of the algorithm, we restrict our derivations to the 1D problem of scattering by an isotropic point (or 0D) `metasurface'. This is sufficient to capture the gist of the GSTC implementation and straightforwardly extends to the case of 2D/3D (1D/2D metasurface) problems and bianisotropic metasurfaces, as detailed in~\cite{GSTC_FDTD}. However, the illustrative example will deal with a 2D metasurface. Furthermore, we assume here that the metasurface is nondispersive, the case of dispersion being treated in Sec.~\ref{sec:FDTD_freq_disp}.

\begin{figure}[!ht]
\centering
\begin{subfigure}{1\columnwidth}
  \centering
  \includegraphics[width=1\columnwidth]{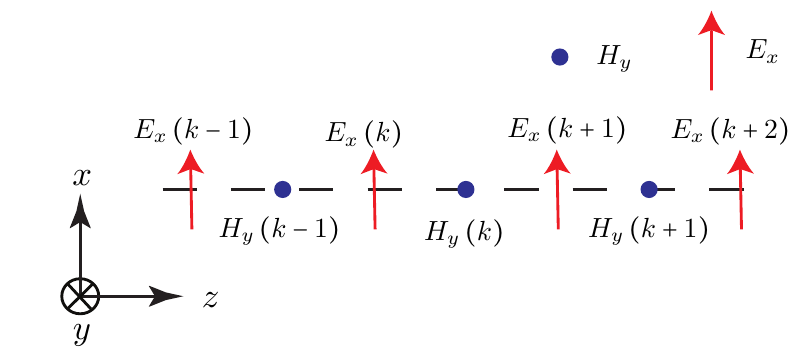}
  \caption{} \label{Fig:1DFDTD_A}
\end{subfigure}
\begin{subfigure}{1\columnwidth}
  \centering
  \includegraphics[width=1\columnwidth]{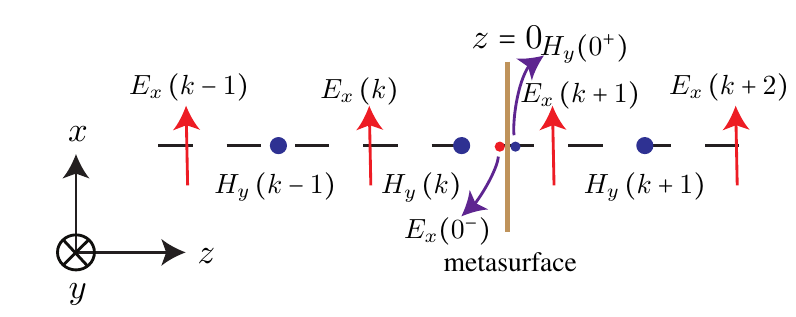}
  \caption{} \label{Fig:1DFDTD_B}
\end{subfigure}
\caption{Integration of the metasurface within the staggered FDTD Yee grid for a 1D (0D or point metasurface) problem with propagation along the $z$-direction. (a)~Conventional grid used in the analysis of non-zero thickness structures. (b)~Positioning of the metasurface between two grid nodes at $z=0$ and introduction of surrounding virtual nodes at $z=0^\pm$.}
\label{Fig:1DFDTD}
\end{figure}

In our 1D problem, we assume propagation in the $z$-direction with nonzero field components $E_z$ and $H_y$, as shown in Fig.~\ref{Fig:1DFDTD_A}. In this case, the GSTCs~\eqref{eq:Freq_GSTC}, expressed in the time-domain (replacing $j\omega$ by $d/dt$) as required here, reduce to
\begin{subequations}\label{TGSTC}
  \begin{equation}\label{TGSTC1}
    -\Delta H_y=\varepsilon_0\frac{d\left[\chi_\textrm{ee}^{xx}E_{x,\textrm{av}} \right]}{dt},
  \end{equation}
  \begin{equation}\label{TGSTC2}
    -\Delta E_x=\mu_0\frac{d\left[\chi_\textrm{mm}^{yy}H_{y,\textrm{av}} \right]}{dt}.
  \end{equation}
\end{subequations}

As in the FDFD case (Sec.~\ref{sec:GSTC-FDFD}), and for the same reasons, the metasurface cannot placed on the grid nodes, and is therefore positioned in between nodes, as shown in Fig.~\ref{Fig:1DFDTD_B}. However, beyond this similarity, there is a fundamental difference between the FDFD and the FDTD metasurface problems. The FDFD algorithm solves the steady-state problem in one shot, via the inversion of a matrix describing the entire computational domain [Eq.~\eqref{eq:FDFD_matrix} and Footnote~\ref{fn:FDTD_matrix_syst}]. In contrast, the FDTD scheme is a march-on-time algorithm, where care must be taken to ensure that the update equations are causal, i.e. that the update fields, at the present time step, do not depend on any quantity at future time steps.

The conventional FDTD update equation for $E_x^n \left(k+1\right)$~\cite{Susan_FDTD} just beyond the metasurface [Fig.~\ref{Fig:1DFDTD_B}] reads
\begin{align}\label{Eyeqstep1}
  E_x^n \left(k+1\right)=&E_x^{n-1}\left(k+1\right)\\\notag
  &+\frac{\Delta t}{\varepsilon_0\Delta z} \left[ H_y^{n-\frac{1}{2}}\left(k+1 \right) -H_y^{n-\frac{1}{2}}\left( k\right) \right].
\end{align}
Since $H_y^{n-\frac{1}{2}}\left( k\right)$ and $E_x^n(k+1)$ are positioned on either side of the metasurface discontinuity, Eq.~\eqref{Eyeqstep1} is not appropriate, for it would fail to account for the presence of the metasurface.

As in the GSTC-FDFD scheme, one may then use~\eqref{TGSTC2} to update $E_x^n \left(k+1\right)$. This yields, after discretization,
\begin{align}\label{GSTC_Ex_disc}
      E_x^n(k+1)=&E_x^n(k)\\\notag&-\frac{\mu_0}{\Delta t}[(\chi_\textrm{mm}^{yy}H_{y,\textrm{av}})^{n+\frac{1}{2}}+(\chi_\textrm{mm}^{yy}H_{y,\textrm{av}})^{n-\frac{1}{2}}],
\end{align}
where $E_x^n(k+1)$ depends on $H_{y,\textrm{av}}^{n+\frac{1}{2}}$, which is at a future time step and is hence unavailable. Thus, this equation violates the aforementioned causality condition and cannot be used \emph{in replacement} of the FDTD equation\footnote{In contrast, in the FDFD scheme, the discretized GSTCs [Eqs.~\eqref{TE_GSTC2_discret} and~\eqref{TE_GSTC1_discret}] could be used in replacement of the corresponding FDFD equations}.

This problem can be resolved by introducing \emph{virtual electric and magnetic nodes}, infinitesimally close to the metasurface, on either side of it, as shown in Fig.~\ref{Fig:1DFDTD_B}~\cite{GSTC_FDTD}. With this strategy, using the magnetic virtual node, at $z=0^+$, Eq.~\eqref{Eyeqstep1} may be approximated\footnote{As in the FDFD case, the FDTD computational metasurface is thus not of exactly zero thickness. Its overall thickness, with~\eqref{Eyeqstep2} and~\eqref{Hxeqstep2}, is $\Delta z$, which corresponds to the mesh resolution. However, this still leads to efficient computation, for the same reason as that explained in the FDFD case (see Footnote~\ref{fn:approx}).} as
\begin{align}\label{Eyeqstep2}
  E_x^n \left(k+1\right)=&E_x^{n-1}\left(k+1\right)\\\notag
  &+\frac{\Delta t}{\varepsilon_0\Delta z} \left[ H_y^{n-\frac{1}{2}}\left(k+1 \right) -H_y^{n-\frac{1}{2}}\left( 0^+\right) \right].
\end{align}
In this relation, $H_y^{n-\frac{1}{2}}\left( 0^+\right)$ can be obtained from the time-domain GSTC equation~\eqref{TGSTC1}, which yields, after discretization,
\begin{subequations}\label{Ey_GSTC_disc}
  \begin{align}\label{Eyeqstep3}
  H_y^{n-\frac{1}{2}}\left( 0^+\right)&=H_y^{n-\frac{1}{2}}\left(k\right)\\\notag&-\frac{\varepsilon_0}{\Delta t}[\left(\chi_\textrm{ee}^{xx}E_{x,\textrm{av}} \right)^n-\left( \chi_\textrm{ee}^{xx}E_{x,\textrm{av}} \right)^{n-1}],
\end{align}
where
\begin{equation}\label{Eyeqstep3_b}
  E_{x,\textrm{av}}^n=\frac{E_x^n(k+1)+E_x^n(k)}{2}.
\end{equation}
\end{subequations}
Substituting this equation into~\eqref{Eyeqstep2} yields the \emph{causal} update equation for $E^n_x(k+1)$
\begin{subequations}
\label{Eyeqstep4}
\begin{align}\label{Eyeqstep4a}
  E_x^n \left(k+1\right)&A_\textrm{ee}^{xx,n}=E_x^{n-1}\left(k+1\right)A_\textrm{ee}^{xx,n-1}\\\notag
  &+\frac{\Delta t}{\varepsilon_0\Delta z} \left[ H_y^{n-\frac{1}{2}}\left(k+1 \right) -H_y^{n-\frac{1}{2}}\left(k\right) \right]\\\notag
  &+\frac{\chi_\textrm{ee}^{xx,n}}{2\Delta z}E_x^n\left(k \right)-\frac{\chi_\textrm{ee}^{xx,n-1}}{2\Delta z}E_x^{n-1}\left(k \right),
\end{align}
where
\begin{equation}
A_\textrm{ee}^{xx,n}=1-\frac{\chi_\textrm{ee}^{xx,n}}{2\Delta z}.
\end{equation}
\end{subequations}

The conventional FDTD update equation for $H_y^{n+\frac{1}{2}}(k)$~\cite{Susan_FDTD} just before the metasurface [Fig.~\ref{Fig:1DFDTD_B}] reads
 \begin{equation}\label{Hxeqstep0}
   H_y^{n+\frac{1}{2}}\left(k\right)=H_y^{n-\frac{1}{2}}\left(k\right)+\frac{\Delta t}{\mu_0\Delta z}\left[E_x^n\left(k+1\right)-E_x^n\left(k\right) \right],
 \end{equation}
where $E_x^n\left(k+1\right)$ and $E_x^n\left(k\right)$ are positioned either side of the metasurface. Using this time the electric virtual node, at $z=0^-$, one obtains upon substituting $E_x^n(0^-)$ into~\eqref{Hxeqstep0},
 \begin{equation}\label{Hxeqstep1}
   H_y^{n+\frac{1}{2}}\left(k\right)=H_y^{n-\frac{1}{2}}\left(k\right)+\frac{\Delta t}{\mu_0\Delta z}\left[E_x^n\left(0^-\right)-E_x^n\left(k\right) \right].
 \end{equation}
In this relation, $E_x^n(0^-)$ is obtained from~\eqref{TGSTC2}, which reads, after discretization,
\begin{subequations}\label{Hx_GSTC_disc}
  \begin{align}\label{Hxeqstep2}
  E_x^n\left( 0^- \right)&=E_x^n\left(k+1 \right)\\\notag  &+\frac{\mu_0}{\Delta t}\left[ \left( \chi_\textrm{mm}^{yy}H_{y,\textrm{av}}\right)^{n+\frac{1}{2}} - \left(\chi_\textrm{mm}^{yy}H_{y,\textrm{av}}\right)^{n-\frac{1}{2}} \right],
  \end{align}
  where
  \begin{equation}\label{Hxeqstep2_b}
   H_{y,\textrm{av}}^{n+\frac{1}{2}}=\frac{H_y^{n+\frac{1}{2}}(k+1)+H_{y}^{n+\frac{1}{2}}(k)}{2}.
  \end{equation}
\end{subequations}
Substituting this noncausal relation into~\eqref{Hxeqstep1} yields the causal update equation for $H_y^{n+\frac{1}{2}}$
\begin{subequations}\label{Hxeqstep3}
   \begin{align}\label{Hxeqstep3a}
   H_y^{n+\frac{1}{2}}&\left(k\right)A_\textrm{mm}^{yy,n+\frac{1}{2}}=H_y^{n-\frac{1}{2}}\left(k\right)A_\textrm{mm}^{yy,n-\frac{1}{2}}\\\notag
   &+\frac{\Delta t}{\mu_0\Delta z}\left[E_x^n\left(k+1\right)-E_x^n\left(k\right) \right]\\\notag &+\frac{\chi_\textrm{mm}^{yy,n+\frac{1}{2}}}{2\Delta z}H_y^{n+\frac{1}{2}}\left(k+1\right)- \frac{\chi_\textrm{mm}^{yy,n-\frac{1}{2}}}{2\Delta z}H_y^{n-\frac{1}{2}}\left(k+1\right).
 \end{align}
 where
\begin{equation}
A_\textrm{mm}^{yy,n}=1-\frac{\chi_\textrm{mm}^{yy,n}}{2\Delta z}.
\end{equation}
\end{subequations}

Comparing~\eqref{Eyeqstep4} and~\eqref{Hxeqstep3} with~\eqref{Eyeqstep1} and~\eqref{Hxeqstep0}, respectively, reveals that the GSTC-FDTD update equations are simply extensions of the conventional FDTD update equations. They can therefore be very easily integrated in a conventional FDTD code at the location(s) of the metasurface(s).

Particularly, in the limit case where the metasurface vanishes, i.e. $\chi=0$, the last two terms disappear and $A_\textrm{ee}^{xx,n}=A_\textrm{mm}^{yy,n}=1$ in~\eqref{Eyeqstep4} and~\eqref{Hxeqstep3}, which immediately leads to~\eqref{Eyeqstep1} and~\eqref{Hxeqstep0}. Moreover, in the next limit case of a time-invariant metasurface, $\chi_\textrm{mm}^{yy,n}=\chi_\textrm{mm}^{yy}$, and therefore $A_\textrm{mm}^{yy,n}=A_\textrm{mm}^{yy}=1-\frac{\chi_\textrm{mm}^{yy}}{2\Delta z}=\mu_r$ in~\eqref{Hxeqstep3}, which upon division by $\mu_r$ reduces to
\begin{equation}\label{Hxeqstep4}
\begin{split}
   H_y^{n+\frac{1}{2}}(k)&=H_y^{n-\frac{1}{2}}\left(k\right)\\
   &+\frac{\Delta t}{\mu_0\mu_\text{r}\Delta z}\left[E_x^n\left(k+1\right)-E_x^n\left(k\right) \right]\\ &+\frac{\chi_\textrm{mm}^{yy}}{2\mu_\text{r}\Delta z}\left[H_y^{n+\frac{1}{2}}\left(k+1\right)- H_y^{n-\frac{1}{2}}\left(k+1\right)\right].
\end{split}
\end{equation}
In this equation, the metasurface information is distributed only between the two terms with square brackets\footnote{Interestingly, without the second term with square brackets, this equation is identical to the conventional FDTD equation~\eqref{Hxeqstep0} for a medium of relative permeability $\mu_\text{r}=1-\frac{\chi_\textrm{mm}^{yy}}{2\Delta z}$, which corresponds to the volume-diluted version of the metasurface surface susceptibility over the distance $2\Delta z$.}, while in the general time-varying case, it is distributed among \emph{all} the terms of the equation, as may be seen for instance by dividing~\eqref{Hxeqstep3} by $A_\textrm{mm}^{yy,n+\frac{1}{2}}$. Similar considerations naturally hold for the time-invariant version of~\eqref{Eyeqstep4} and its time-variant generalization~\ref{Eyeqstep4}].

The GSTC-FDTD scheme presented in this section does not include the possibility to account for specific (causal) metasurface dispersions (e.g. Lorentz, Drude, Debye). However, it automatically accounts for the dispersion form $\tilde{\chi}(\omega)=\frac{\alpha}{j\omega}$, where $\alpha$ is a constant, which may represent a limited-band approximation of several (causal) dispersive responses~\cite{Cloud_Rothwell}\footnote{Particularly, the inverse Fourier transform, or impulse response (not to be confused with the time variation of the time-varying metasurface), of the corresponding permittivity (or corresponding quantity) function is real, since $-j\frac{\alpha}{\omega}$ is an odd function of $\omega$ and the corresponding real part of the permittivity, as a constant, is an even function of $\omega$.}. Indeed, for a monoisotropic metasurface with such dispersion, which reduces to the susceptibilities
$\tilde{\chi}_\textrm{ee}^{xx}=\tilde{\chi}_\textrm{mm}^{yy}=\frac{\alpha}{j\omega}$,
Eqs.~\eqref{eq:Freq_GSTC} reduce to
\begin{subequations}\label{Disp_chi}
  \begin{align}\label{Disp_chi1}
    -\Delta H_y&=\varepsilon_0\alpha E_{x,\textrm{av}},\\\label{Disp_chi2}
    -\Delta E_x&=\mu_0\alpha H_{y,\textrm{av}},
  \end{align}
\end{subequations}
where the tildes have been removed since the spectral relations are the same as their time counterparts in the absence of frequency dependence. Properly discretizing these relations provides for the required fields at the virtual nodes
\begin{subequations}\label{Disp_chi_disc}
  \begin{align}\label{Disp_chi_disc1}
  &H_y^{n-\frac{1}{2}}(0^+)=H_y^{n-\frac{1}{2}}(k)-\frac{\varepsilon_0\alpha}{4} \\\notag& \left[ E_x^n(k)+E_x^n(k+1)+E_x^{n-1}(k)+E_x^{n-1}(k+1)\right],\\\label{Disp_chi_disc2}
  &E_x^n(0^-)=E_x^n(k+1)+\frac{\mu_0\alpha}{4} \\\notag& \left[ H_y^{n-\frac{1}{2}}(k)+H_y^{n+\frac{1}{2}}(k)+H_y^{n-\frac{1}{2}}(k+1)+H_y^{n+\frac{1}{2}}(k+1)\right],
  \end{align}
\end{subequations}
which are to be used in replacement of~\eqref{Ey_GSTC_disc} and~\eqref{Hx_GSTC_disc}, respectively. Since they are frequency-independent, are compatible with time-domain equations and may then be straightforwardly inserted into~\eqref{Eyeqstep2} and~\eqref{Hxeqstep1}, respectively, to provide the corresponding FDTD equations.

\begin{figure}[!ht]
\centering
\begin{subfigure}{1\columnwidth}
  \centering
  \includegraphics[width=1\columnwidth]{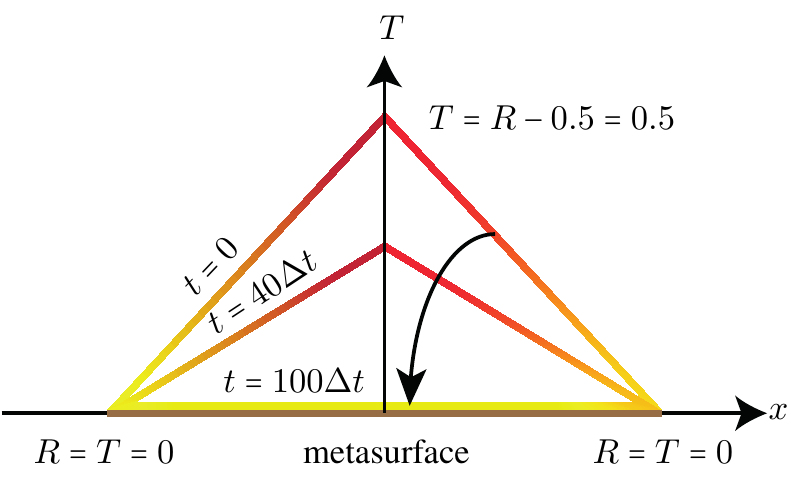}
  \caption{} \label{Fig:st_profile_A}
\end{subfigure}

\begin{subfigure}{1\columnwidth}
  \centering
  \includegraphics[width=1\columnwidth]{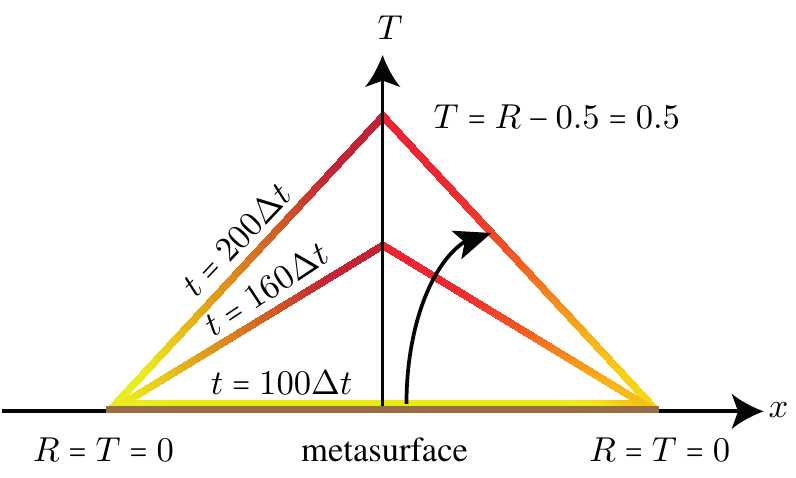}
\caption{} \label{Fig:st_profile_B}
\end{subfigure}
\caption{Harmonically time-varying 1D metasurface (2D problem) susceptibility with triangular spatial profile. (a)~First half of the periodic cycle. (b)~Second half of the periodic cycle.}
\label{Fig:st_profile}
\end{figure}

An example of space-time varying dispersive metasurface with the aforementioned dispersion\footnote{Specifically, $\alpha=1$ and $\alpha=3$ when $T=R=0$ and $T=R-0.5=0.5$, respectively (see~\cite{GSTC_FDTD} for details).} is shown is Fig.~\ref{Fig:st_profile}. Initially, at $t=0$, the metasurface absorbs all the incident wave at its edges and transmits of the field amplitude at its center, with linear variation from the edges to the center (triangular profile). This spatial susceptibility function varies harmonically in time, period $t=200\Delta t$, pass through a full absorber state at $t=100\Delta t$.

\begin{figure}[!ht]
\centering
\includegraphics[width=1\columnwidth]{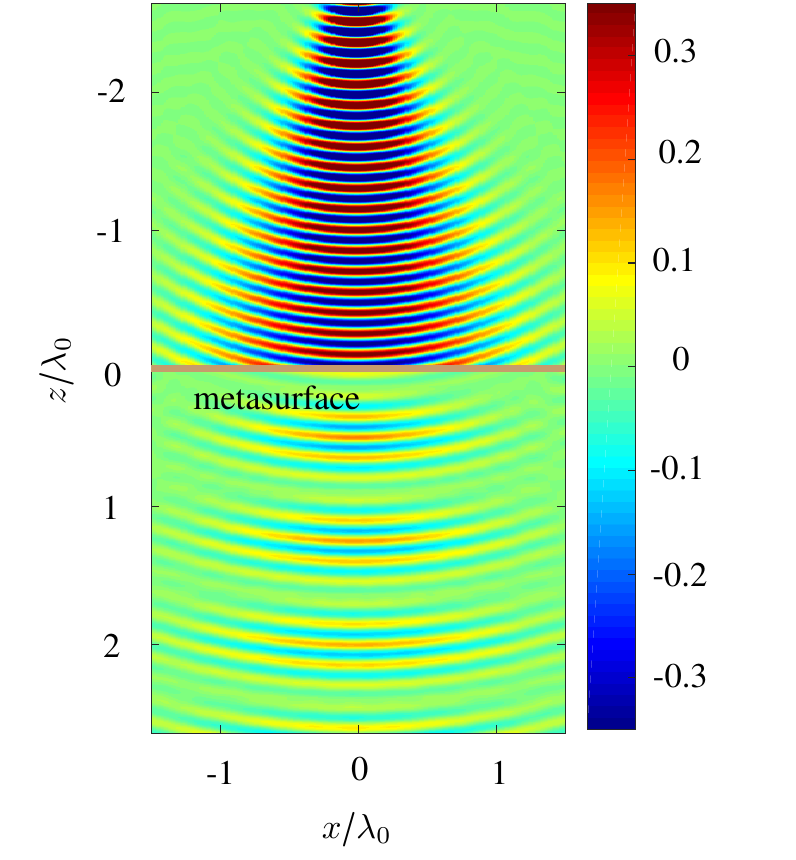}
  \caption{FDTD-simulated electric field ($E_x$) in space for the space-time varying metasurface of Fig.~\ref{Fig:st_profile}.} \label{Fig:FDTD2DAbsorber}
\end{figure}

\begin{figure}[!ht]
\centering
\includegraphics[width=1\columnwidth]{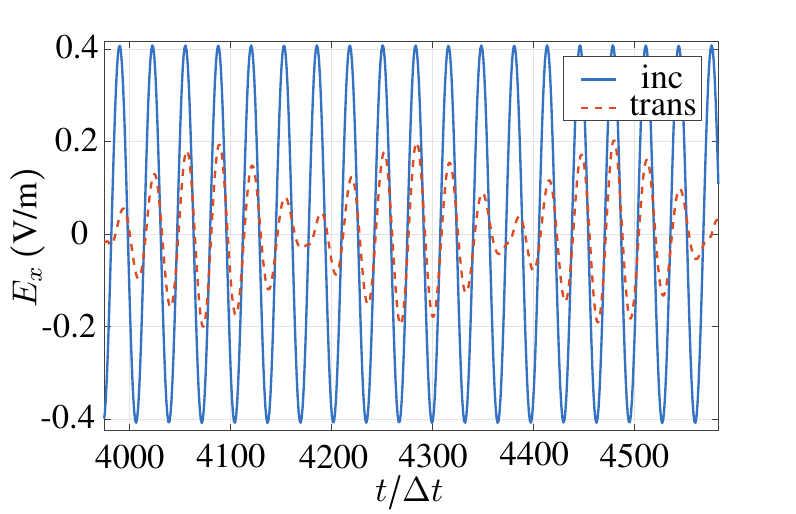}
\caption{FDTD-simulated field in time, at $(x,z)=(0,0^+)$ (incident wave) and $(x,z)=(0,0^-)$ (transmitted wave) in Fig.~\ref{Fig:FDTD2DAbsorber} for the space-time varying metasurface of Fig.~\ref{Fig:st_profile}.}\label{Fig:FT}
\end{figure}

Figures~\ref{Fig:FDTD2DAbsorber} and~\ref{Fig:FT} show GSTC-FDTD results for the space-time varying metasurface in Fig.~\ref{Fig:st_profile}. In Fig.~\ref{Fig:FDTD2DAbsorber}, the highest and lowest amplitudes of the transmitted field corresponds to the times when the metasurface was maximally transmitting and fully absorbing, respectively. The reflected field is zero, as specified, as may be inferred from the absence of standing wave pattern on the incident side ($z<0$) of the metasurface. In Fig.~\ref{Fig:FT}, the steady-state time variation of the field in the incident and transmitted field regions confirm the harmonic variations of the field and the absence of reflection.

\subsection{Extension to Arbitrary Frequency Dispersions}\label{sec:FDTD_freq_disp}

As mentioned above, the GSTC-FDTD scheme described in Sec.~\ref{sec:FDTD}, while otherwise very general and flexible, does not allow the specification of arbitrary frequency-dispersive susceptibilities, except for the trivial case of zero dispersion [$\tilde{\chi}\neq\tilde{\chi}(\omega)$] and the particular case of the dispersion form $\tilde{\chi}(\omega)=\frac{\alpha}{j\omega}$. However, the integration of arbitrary dispersion is not a fundamental difficulty and may be done using conventional techniques~\cite{Susan_FDTD}.

This was first done for time-invariant and time-varying metasurfaces in~\cite{Shulabh_FDTD3} and~\cite{Shulabh_FDTD2}, respectively, which implemented the GSTCs with Lorentz dispersion in terms of an LC-circuit model. This model requires drastic transformation in a conventional FDTD code and precludes the introduction of scattering objects in the computational domain. These limitations were obviated in~\cite{Shulabh_FDTD1}, which implemented the GSTCs in a similar fashion as~\cite{GSTC_FDFD}, and used Auxiliary Differential Equations (ADEs) to account for the Lorentzian dispersion of an isotropic metasurface. However, the ADE imlementation in this paper is such  that it requires the resolution of a matrix system at each time step.

This issue was removed in~\cite{Usef_dispersive}, which presented a general GSTC-FDTD scheme for bianisotropic metasurfaces with Lorentzian dispersion. In this scheme, dispersion is efficiently treated by ADEs, using the bianisotropic polarization density auxiliary functions $P_\textrm{ee}, P_\textrm{em}, M_\textrm{mm}$ and $M_\textrm{me}$, and is straightforwardly extendible to other dispersion forms via ADE. The reader is referred to~\cite{Usef_dispersive} for details.

\section{Conclusion and Prospects}\label{sec:conclusion}

The design of a metasurface requires a holistic approach involving synergistic synthesis and analysis operations. In this approach, the metasurface is best modeled as a zero-thickness sheet characterized by a surface susceptibility tensor function and described by GSTCs. This has been demonstrated in both frequency-domain (FDFD, SD-IE, FEM) and time-domain (FDTD) computational techniques, where the GSTCs, integrable within the core algorithms, lead to the most powerful computational analysis of general bianisotropic space-varying and time-varying metasurfaces.

The presented GSTC computational analysis of metasurfaces represents the solution to the very canonical problem of deeply subwavelength electromagnetic sheets. Beyond metasurfaces, it may also apply to 2D electron gas (2DEGs), emerging 2D materials (e.g. graphene, MO$_2$, black phosphorous) and other 2D composite structures.

At this point, several types of generalizations - ultimately leading to the accommodation of general 3D bianisotropic, space-time varying and nonlinear metasurfaces in arbitrary environments - can be straightforwardly worked out and implemented in commercial softwares. The different GSTC-based techniques can also be extended to analyze nonplanar (cylindrical, spherical, etc.) metasurfaces~\cite{Safgri_Cylindr_ms,Xiao_Spherical_ms}.

A still unsolved non-trivial problem, however, that is the extension to metasurfaces including normal polarization densities, i.e. 3D tensors replacing the 2D tensors in~\eqref{eq:Freq_GSTC}, which will allow the analysis of more sophisticated and powerful metasurfaces.

Finally, the extension of the proposed concepts to 2D multiphysics structures, accounting for latest technological developments, will most likely represent a substantial part of future research on the computational analysis of metasurfaces.

%

\bibliographystyle{IEEEtran}
\bibliography{JMMCT_METASURFACE_INVITED_Vahabzadeh}


\end{document}